\documentclass[lettersize,journal]{IEEEtran}
\usepackage{amsmath,amsfonts}
\usepackage{algorithmic}
\usepackage{algorithm}
\usepackage{array}
\usepackage[caption=false,font=normalsize,labelfont=sf,textfont=sf]{subfig}
\usepackage{textcomp}
\usepackage{stfloats}
\usepackage{url}
\usepackage{verbatim}
\usepackage{graphicx}
\usepackage{cite}
\usepackage{amssymb}
\usepackage{booktabs} 
\usepackage{siunitx}
\usepackage{xcolor}
\begin{document}

\title{Perceptual Characterization of \\ Scan-induced Artifacts in Dynamic 3D Meshes}

\author{Huong Hoang, Truong Nguyen, Pamela Cosman \\
Department of Electrical and Computer Engineering, University of California, San Diego, CA, USA 92093
\thanks{This work was supported in part by the Center for Wireless Communications, University of California, San Diego. }}

\maketitle

\begin{abstract}
Dynamic meshes reconstructed from real-world scenes often exhibit scan-induced artifacts such as surface jitter and missing geometry, which strongly impact perceived visual quality. Objective assessment remains challenging due to the absence of artifact-free ground truth and limited perceptual understanding of artifact interactions. This paper presents a systematic and controlled perceptual study of scan-induced artifacts in dynamic meshes, focusing on surface jitter and holes. To enable systematic analysis, we introduce a physically grounded artifact synthesis framework in which holes are synthesized using a reconstruction uncertainty model derived from multi-view geometry. 
The study reveals asymmetric cross-artifact interactions, non-additive integration, and motion-dependent temporal masking. Based on these findings, we propose a compact, perceptually grounded no-reference quality assessment framework that encodes the observed perceptual mechanisms. 
\end{abstract}


\section{Introduction}

\begin{figure}[h]
    \centering
    \includegraphics[width=0.82\linewidth]{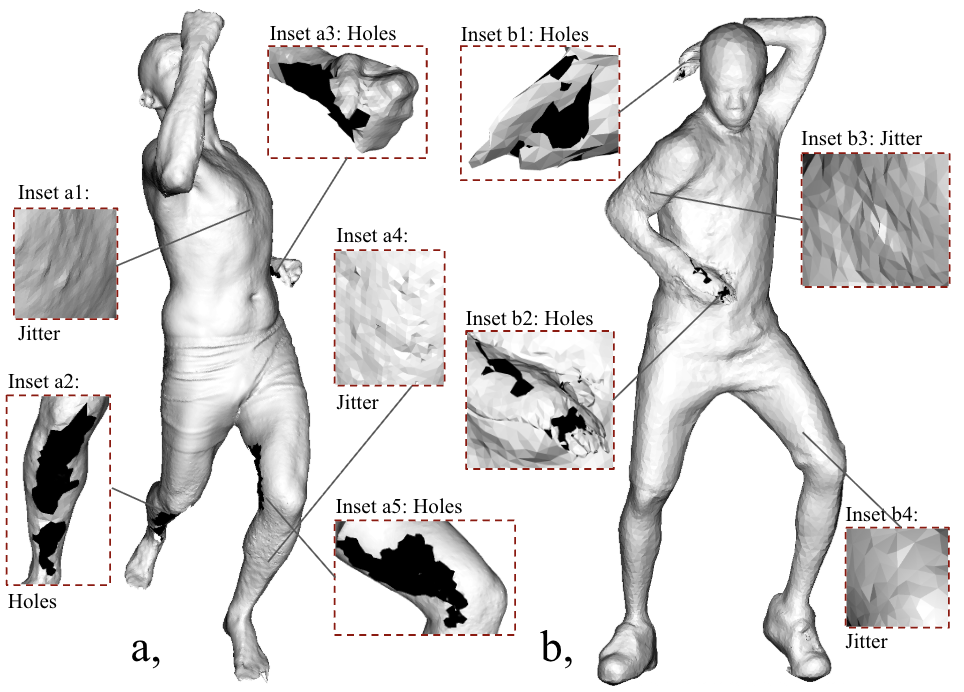}
    \caption{Examples of scan-induced artifacts in dynamic mesh datasets. \textbf{Jitter} produces high-frequency geometric surface instability over otherwise smooth regions, whereas \textbf{holes} appear as localized missing regions with sharp boundaries.
        \textit{(a)} MANO dataset \cite{MANO_dataset}, model B08.
        \textit{(b)} Volograms dataset  \cite{volograms_dataset}, model Levi.}
    \label{fig:artifacts-vis}
\end{figure}

Dynamic mesh reconstruction from real-world scenes is increasingly used in immersive telepresence, virtual reality, and free-viewpoint video. Despite advances in multi-view capture and reconstruction algorithms, dynamic meshes reconstructed from real objects remain affected by scan-induced artifacts, such as surface jitter and missing geometry/holes (Fig. \ref{fig:artifacts-vis}). These artifacts arise from fundamental limitations in camera coverage, correspondence estimation, and triangulation, and persist even in high-quality capture pipelines.

Assessing the perceptual quality of such dynamic meshes remains an open challenge. Unlike synthetic animations, reconstructions of real scenes lack an artifact-free ground truth; reference sequences are themselves reconstructed and therefore contain residual noise and temporal inconsistencies. As a result, reference-based metrics may penalize deviations that are perceptually beneficial. This limitation motivates perceptual studies and no-reference quality assessment approaches that do not assume access to a perfect reference.

Most existing objective metrics for 3D geometry focus on static meshes or assume reference-based evaluation. While recent work has extended quality assessment to dynamic meshes, these efforts primarily address compression artifacts and rely on reference sequences. More fundamentally, existing approaches typically assume that multiple distortions combine additively or independently. However, perceptual studies in image and video quality assessment consistently show that perceived quality does not follow simple additive rules when multiple impairments are present. Whether similar interaction behavior exists for reconstruction artifacts in dynamic meshes, and how such interactions manifest perceptually, remains largely unexplored.

In this paper, we present a systematic perceptual study of reconstruction artifacts in dynamic meshes, focusing on surface jitter and missing geometry. To enable precise experimental control, we introduce a novel artifact synthesis framework in which holes are synthesized using a physically grounded uncertainty model derived from multi-view reconstruction theory. This allows realistic hole generation without relying on heuristic removal or algorithm-specific failures.

Using this framework, we conduct a subjective experiment that systematically varies artifact severity, artifact combinations, and motion characteristics. The results reveal several previously unreported perceptual phenomena: (i) asymmetric cross-artifact interactions, where holes exaggerate perceived jitter (amplification effect) while jitter suppresses perceived hole visibility (masking effect); (ii) non-additive integration of multiple artifacts; and (iii) motion-dependent temporal masking, in which fast motion selectively reduces sensitivity to jitter while leaving hole perception largely unaffected.

Building on these findings, we propose a no-reference quality assessment framework that encodes experimentally observed perceptual mechanisms using a small number of interpretable parameters. 

Overall, the main contributions of this work are:
\begin{itemize}
    \item A physically grounded artifact synthesis pipeline enabling controlled generation of scan-induced artifacts, including the first synthesis of missing geometry based on multi-view reconstruction uncertainty for controlled perceptual evaluation.
    \item A comprehensive perceptual study resulting in the \textit{DynaScan} dataset, revealing asymmetric artifact interactions, non-additive integration, motion-dependent masking, and observer-specific dominance patterns in dynamic mesh perception.
    \item A compact, perceptually grounded no-reference quality assessment framework demonstrating how these perceptual findings can inform objective modeling.
\end{itemize}

Together, these contributions advance the understanding of reconstruction artifact perception in dynamic meshes and provide a principled foundation for future perceptual modeling and quality assessment.

\section{Related Work}
\subsection{Perceptual Study and Assessment of Static 3D Geometry}

Objective quality assessment of static 3D meshes has been extensively studied. Early metrics focused on purely geometric deviation measures, such as vertex-to-surface distances and global surface deviation, such as Metro \cite{Cignoni1998} and GL \cite{Karni2000}. These methods quantify numerical error but do not explicitly model human perception.

Subsequent work introduced perceptually motivated geometric metrics, validated through controlled subjective experiments. Notable examples include 3DWPM \cite{4066988}, MSDM \cite{Lavou2006}, MSDM2 \cite{Lavou2011}, FMPD \cite{Wang2012}, and DAME \cite{Va2012}. These models incorporate curvature, structural distortion, or local saliency considerations, demonstrating significantly improved correlation with human opinion scores. In parallel, psychophysical studies such as \cite{10.1145/383259.383283} investigated human sensitivity to static geometric distortions without proposing explicit metrics.

Despite their success, these approaches are inherently frame-based and temporally agnostic. They assume static inspection and do not account for temporal integration, temporal masking, or motion-induced perceptual effects. Consequently, they cannot explain quality judgments of dynamic 3D meshes where temporal instability plays a dominant role.

\subsection{Perceptual Study and Assessment of Dynamic 3D Meshes}
Compared to static meshes, perceptual quality assessment of dynamic 3D meshes remains less explored. Early work, such as STED \cite{Vasa2011}, computes the spatiotemporal edge difference to estimate the quality of dynamic meshes, while \cite{Torkhani2015} conducted subjective and objective evaluations, highlighting the perceptual importance of temporal distortions. However, most existing dynamic mesh metrics assume consistent mesh topology across frames, enabling vertex correspondence and temporal tracking. This assumption breaks down for real-world dynamic reconstructions from multi-view capture systems, where topology may vary due to tracking failure or occlusion.

More recent studies have attempted to bridge this gap. Yildiz et al. \cite{Yildiz2017} extended image-space sensitivity models to dynamic 3D representations. Yang et al. \cite{Yang2025} provided a comprehensive subjective and objective evaluation of existing dynamic mesh metrics. Nevertheless, these approaches operate within a reference-based paradigm, assuming access to a pristine ground-truth mesh sequence.
In practice, however, ground-truth dynamic meshes, especially those reconstructed from real multi-view systems, already contain residual artifacts. This limits the interpretability and reliability of purely reference-based evaluation for real-world capture scenarios.

\subsection{Perceptual Study of Scan-induced Arifacts}
Existing perceptual studies on dynamic meshes \cite{Torkhani2015, Yildiz2017, Yang2025} primarily focus on controlled geometric distortions such as simplification, quantization, or compression noise.
These distortions are typically introduced through explicit parametric operations, enabling systematic control over their magnitude and statistical properties.
In contrast, scan-induced artifacts such as holes originate from multi-view reconstruction processes and are driven by visibility inconsistencies, triangulation uncertainty, and reconstruction confidence. Rather than being explicitly parameterized, their spatial structure reflects the underlying imaging geometry and reconstruction dynamics.

A key obstacle is stimulus generation. Unlike additive geometric noise, realistic missing geometry cannot be generated by random vertex removal. Their structure depends on visibility, triangulation uncertainty, and surface completeness constraints. As a result, perceptually controlled experiments on scan-induced artifacts remain scarce. 
This gap motivates a principled framework for generating and studying scan-induced artifacts under controlled conditions.

\subsection{Reconstruction Uncertainty and Perceptual Relevance}
Geometric uncertainty in multi-view 3D reconstruction has been extensively analyzed in computer vision and photogrammetry. Concepts like covariance propagation and uncertainty ellipsoids are well established in 3D reconstruction literature. Classical treatments describe how image noise propagates to 3D point estimates through Jacobian-based covariance analysis \cite{Hartley2004}, explaining the anisotropic, view-dependent, and baseline-sensitive nature of reconstruction errors.

However, this body of work is mainly used for numerical accuracy and has never been used to synthesize perceptually controlled stimuli or to study how scan-induced artifacts are perceived. In this work, reconstruction uncertainty is repurposed as a generative mechanism to create perceptually controlled, physically plausible scan-induced artifacts, bridging geometric estimation theory and perceptual quality modeling.

\subsection{Integration of Multiple Distortions}
The perceptual integration of multiple distortions has been widely studied in image and video quality assessment. Prior work consistently shows that perceived quality does not follow additive rules when distortions co-occur \cite{Silva2018, 5404314}. Interaction-aware models with multiplicative or non-linear coupling terms significantly outperform additive formulations \cite{5404314}.

In the context of 3D perceptual quality assessment, however, distortion interaction remains largely unexplored. Existing 3D mesh metrics typically assume independence between artifact types and aggregate features additively. To our knowledge, no prior work reports experimentally derived iso-quality contours or explicitly models interaction effects in dynamic 3D scan-induced artifacts.
This represents a significant conceptual gap between 2D video perception theory and 3D dynamic geometry assessment.

\subsection{Motivation for No-Reference Modeling}
Most existing objective metrics for dynamic meshes rely on reference-based evaluation, implicitly assuming that the reference sequence represents artifact-free geometry. In practice, however, reference dynamic meshes are typically reconstructed from real multi-view capture systems and therefore already contain residual scan-induced artifacts. As a result, reference-based metrics may penalize processing methods, such as compression or filtering, that deviate from the reference while improving perceived visual quality.

Moreover, for reconstructions of real objects, an artifact-free ground truth does not exist. This fundamentally limits the applicability of absolute reference-based evaluation and motivates perceptual studies and no-reference quality models that do not assume access to a perfect reference, but instead rely on perceptually meaningful characterization of artifacts.

\section{Artifact Generation Methods}
To study the perceptual impact of scan-induced artifacts in a controlled manner, we synthesize two dominant scan-induced artifacts commonly observed in multi-view capture systems: surface jitter and holes. The two artifacts are generated independently and parametrically, enabling precise control over their severity and combinations while preserving the underlying geometry and motion of the original sequence.

Importantly, while surface jitter is modeled using a generic stochastic process, holes are synthesized using a physically grounded uncertainty model derived from multi-view reconstruction theory. To our knowledge, this is the first perceptual study that explicitly leverages geometric uncertainty theory to synthesize missing geometry in dynamic meshes.

\subsection{Surface Jitter Synthesis}
Surface jitter refers to small, high-frequency perturbations in the positions of reconstructed vertices. Jitter may emerge from fundamental limitations and inconsistencies in data representation and transformation, including noise, discretization, information loss, compression artifacts, and imperfections in data processing pipelines. In reference mesh reconstruction, it is commonly caused by calibration noise, limited camera resolution, correspondence errors, and view-dependent triangulation instability. 

Following common practice in mesh distortion studies, jitter is generated by applying geometrical Gaussian noise directly to the spatial coordinates of mesh vertices \cite{Yang2025}. Let $v_i^t \in \mathbb{R}^3$ denote the position of vertex $i$ at frame $t$. The jittered vertex $\tilde{v}_i^t$ is obtained as:
\begin{equation}
    \tilde{v}_i^t = v_i^t + r_i^t\times Ratio \times minBB^t
\end{equation}

\noindent where $r_i^t$ represents the random noise for $v_i^t$, $Ratio$ controls the jitter severity, and $minBB^t$ denotes the minimum size of frame $t$'s bounding box among three dimensions. 

\subsection{Hole Generation via Reconstruction Uncertainty}\label{sec:hole_gen}
In contrast to surface jitter, holes are not generated heuristically. Instead, they are synthesized based on geometric uncertainty theory, reflecting regions where multi-view reconstruction is intrinsically unreliable due to occlusion and limited baseline.

\subsubsection{Reconstruction Uncertainty Estimation}
To simulate realistic scan-induced artifacts, we model the scanning process by virtually placing multiple cameras around the object in a hemispherical arrangement, mimicking a typical multiview reconstruction setup. Reconstruction uncertainty can be quantified through the 3D covariance matrix of each vertex (i.e., the triangulated point) \cite{Hartley2004}, as:

\begin{equation}
\boldsymbol{\Sigma}_{i} = \mathbf{J}^{-1} \boldsymbol{\Sigma}_{\mathbf{u}} (\mathbf{J}^{-1})^{\top}
\end{equation}

\noindent where $\mathbf{J}$ is the Jacobian of the projection function with respect to vertex $i$, and $\boldsymbol{\Sigma}_{\mathbf{u}}$ models image-space noise. $\mathbf{J}$ can be calculated through the cameras’ intrinsic and extrinsic parameters (e.g., focal length, its position in world axes, and its rotation relative to world axes), along with the vertex position $v_i$. This formulation captures the view-dependent and anisotropic nature of reconstruction uncertainty.

For each vertex, we compute a scalar uncertainty measure from the covariance matrix. Let:
\begin{equation}
\sigma^t_i = \lambda_{\max}(\boldsymbol{\Sigma}^t_{i}),
\end{equation}
\noindent where $\lambda_{\max}$ denotes the largest eigenvalue. $\sigma^t_i$ represents the principal uncertainty magnitude of vertex $i$ at frame $t$. Vertices for which the information matrix is rank-deficient (e.g., visible in fewer than two cameras) are marked as unsupported.

For each face $f$ at frame $t$, with vertex set $\mathcal{V}_f$, we define the face uncertainty magnitude $U^t_f$ as the largest vertex uncertainty magnitude of its belonging vertices:
\begin{equation}
    U^t_f = \max_{v \in \mathcal{V}_f} \sigma^t_v
\end{equation}

This max-based aggregation reflects the fact that a single highly uncertain vertex is sufficient to destabilize the entire face. If any vertex of face $f$ is unsupported, $U^t_f$ is set to NaN, and the face is removed. 

\subsubsection{Surface-Area-Controlled Hole Severity}
Directly thresholding uncertainty values does not yield consistent perceptual severity across sequences due to varying motion, occlusion patterns, and mesh density. To achieve perceptually meaningful and sequence-consistent control, hole severity is defined in terms of average removed surface area.

Let $A_{\mathrm{removed}}(t)$ and $A_{\mathrm{total}}(t)$ denote the removed and total mesh surface area at frame $t$. The average removed area ratio for a mesh sequence of $T$ frames is defined as:

\begin{equation}
    \%A
=
\frac{1}{T}
\sum_{t=1}^{T}
\frac{A_{\mathrm{removed}}(t)}
     {A_{\mathrm{total}}(t)}
\end{equation}

Given a target removal ratio $\%A_{\mathrm{target}}$, we determine a global uncertainty threshold
$\beta$ such that:
\begin{equation}
    \%A(\beta) \approx \%A_{\mathrm{target}}
\end{equation}

To compute $\beta$, we first collect all valid (non-NaN) face uncertainties across all frames:
\begin{equation}
    \mathcal{U}
=
\bigcup_{t=1}^{T}
\left\{
U^t_f
\;\middle|\;
U^t_f \neq \text{NaN}
\right\}
\end{equation}

Faces with unsupported vertices define a lower bound $\%A_{\mathrm{min}}$. If $\%A_{\mathrm{target}} < \%A_{\mathrm{min}}$, no additional faces are removed. Otherwise, valid face uncertainties in $\mathcal{U}$ are sorted in descending order, and faces are accumulated until the target removable surface area is reached. The uncertainty value at the cutoff defines the threshold $\beta$.

In other words, all faces satisfying $U^t_f > \beta$ are removed at frame $t$, resulting in contiguous regions of missing surface that resemble naturally occurring scan holes, whose extent is controlled directly by the target removed surface area ratio.

\section{Perceptual Experiment}
\subsection{Clean Baseline Sequences}
\begin{figure}[h]
    \centering
    \includegraphics[width=0.8\linewidth]{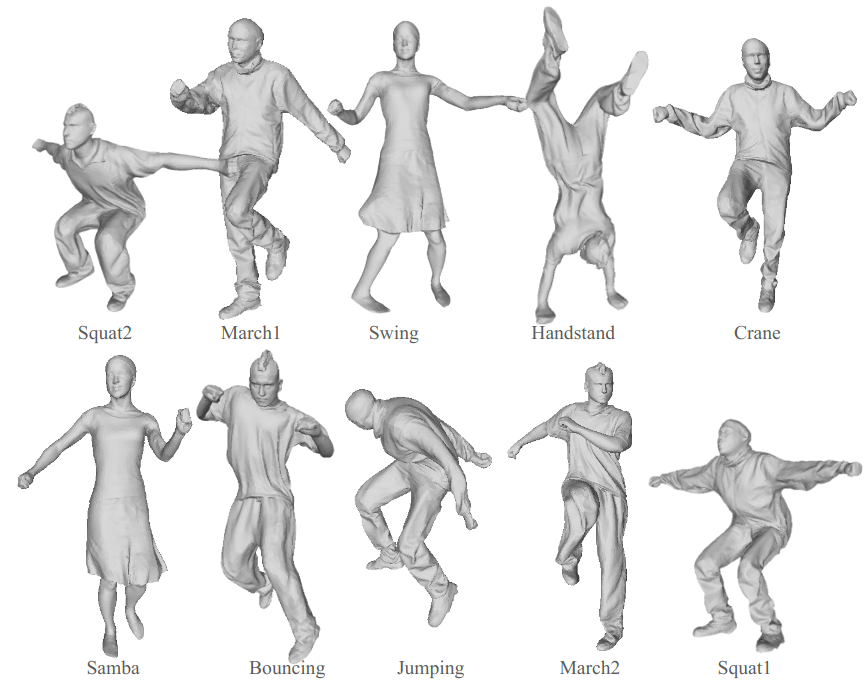}
    \caption{Clean baseline dynamic mesh dataset \cite{vlasic2008articulated}.}
    \label{fig:content-vis}
\end{figure}

All stimuli are derived from a clean dynamic mesh dataset \cite{vlasic2008articulated} (Fig. \ref{fig:content-vis}) of sequences that are high-quality reconstructions with stable topology and minimal visible artifacts, selected to represent a range of human motions and motion intensities.

The dataset consists of ten sequences depicting human motion; eight are used in the perceptual study, while two are reserved for training. Each sequence has a duration between 5 and 8 seconds and is rendered at 30 frames per second.

To characterize content variability, we quantify temporal and spatial complexity using motion and geometry descriptors. Temporal Information (TI) is quantified following the formulation in~\cite{Torkhani2015}. For a sequence of $T$ mesh frames $\mathcal{M}_t = (V_t, F_t)$, let $S_t(v)$ denote the magnitude of the motion vector at vertex $v \in V_t$, computed as the mean amplitude of its forward and backward displacements. TI is defined as
\begin{equation}
\mathrm{TI} = \frac{1}{T} \sum_{t=1}^{T} \frac{1}{|V_t|} \sum_{v \in V_t} S_t(v).
\label{eq:TI}
\end{equation}

To quantify spatial complexity, we define a Spatial Information (SI) measure based on surface curvature variability. Specifically, SI is computed as the temporal average of the variance of the per-vertex Gaussian curvature $K_t(v)$:
\begin{equation}
\mathrm{SI} = \frac{1}{T} \sum_{t=1}^{T} \operatorname{Var}_{v \in V_t}\!\left[ K_t(v) \right].
\end{equation}

\begin{figure}[h]
    \centering
    \includegraphics[width=0.8\linewidth]{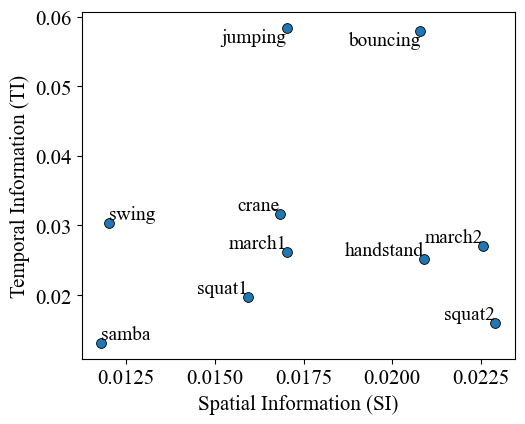}
    \caption{Content characterization in the SI–TI space.}
    \label{fig:content-TI-SI}
\end{figure}

Fig.~\ref{fig:content-TI-SI} shows the distribution of the selected sequences in the SI-TI space, illustrating coverage of diverse spatial and temporal characteristics.

The clean sequences are not treated as artifact-free ground truth; rather, they provide a consistent baseline to which controlled artifact severity is introduced. 
\subsection{Stimulus Construction}
Artifact severity is controlled independently for surface jitter and holes.

\begin{itemize}
    \item Surface jitter severity is determined by the standard deviation of the Gaussian noise applied to vertex positions. Five monotonically increasing levels are selected to span the full perceptual range from imperceptible to strongly annoying jitter.

    \item Hole severity is controlled using the average removed surface area ratio $\%A_{\mathrm{target}}$, defined in Section \ref{sec:hole_gen}. Five target values of $\%A_{\mathrm{target}}$ are chosen to produce perceptually distinct levels of hole visibility, ranging from sparse, barely noticeable holes to large, dominant missing regions.
\end{itemize}

To enable systematic analysis of artifact interactions, stimuli are generated according to three conditions:

\begin{itemize}
    \item Jitter-only: Clean sequences with surface jitter applied at each of the five jitter severity levels, without holes.
    \item Holes-only: Clean sequences with holes synthesized using the uncertainty-based method at each of the five hole severity levels, without jitter.
    \item Combined artifacts: All pairwise combinations of jitter and hole severity levels.
\end{itemize}

This design allows direct comparison between single-artifact and multi-artifact conditions and supports analysis of cross-artifact perceptual effects. In total, the study comprises 280 stimuli. The stimuli are divided into two non-overlapping subsets, each containing four motion sequences, to limit session duration and participant fatigue. To monitor attention and rating reliability, five clean stimuli are inserted at random positions within each session. 

\subsection{Rendering and Presentation}
Some related work used a rotating camera to capture different angles to avoid generating separate videos for multiple viewpoints. However, the rotation path and speed may mask certain artifacts. To avoid that, we keep the camera fixed. Each clip is rendered from two static views (front and back). Each stimulus will display both front and back views simultaneously (split-screen format). 
Similar to \cite{Yang2025}, we use the MPEG 3D renderer to visualize the dynamic sequences at a resolution of 1920×1080. The rendered outputs are then encoded and saved as MP4 videos using FFmpeg. Following \cite{Yang2025}, we apply a constant rate factor of 10 during compression, ensuring smooth playback and visually lossless quality. 
\subsection{Experimental Procedure}
The experiment follows a single-stimulus Absolute Category Rating (ACR) methodology. Each stimulus is presented in a continuous loop, and participants are asked to provide ratings along three dimensions: Jitter annoyance, Hole annoyance, and Overall visual annoyance.

Ratings are collected using a five-point impairment scale:
(1) Imperceptible; (2) Perceptible, but not annoying; (3) Slightly annoying; (4) Annoying; and (5) Very annoying.

Before the main session, participants complete a training phase with example stimuli illustrating the full severity range for each artifact type. These examples are distinct from the test stimuli and are shown to ensure consistent interpretation of the rating scale.

The experiment is conducted in a controlled laboratory environment following ITU-R BT.500 recommendations \cite{ITUR_BT500}. Viewing distance, ambient illumination, and display settings are kept constant across sessions. A total of 46 individuals (ages 18-40; 39 men and 7 women) participated in the study, with 23 participants allocated to each stimulus subset. Each session lasts approximately 45 minutes, including training and scheduled breaks. Participants are encouraged to take breaks as needed, with a mandatory break midway through the session. All participants provided informed consent prior to participation. This study was approved by the Institutional Review Board of UC San Diego (Protocol 210286, Date 4/1/2021).

Post-screening of subjects is performed using correlation-based criteria as specified in ITU-R BT.500 \cite{ITUR_BT500}. One participant whose ratings show insufficient consistency with the group was excluded from further analysis, and data from 45 participants were retained as valid.

\subsection{Data Validity and Rating Behavior}

Before analyzing perceptual interactions, we examine the distribution and consistency of subjective ratings to verify that the experimental design effectively spans the intended perceptual space and that participants engaged meaningfully with the task.

\subsubsection{Rating Range and Stimulus Coverage}
\begin{figure}[h]
    \centering
    \includegraphics[width=\linewidth]{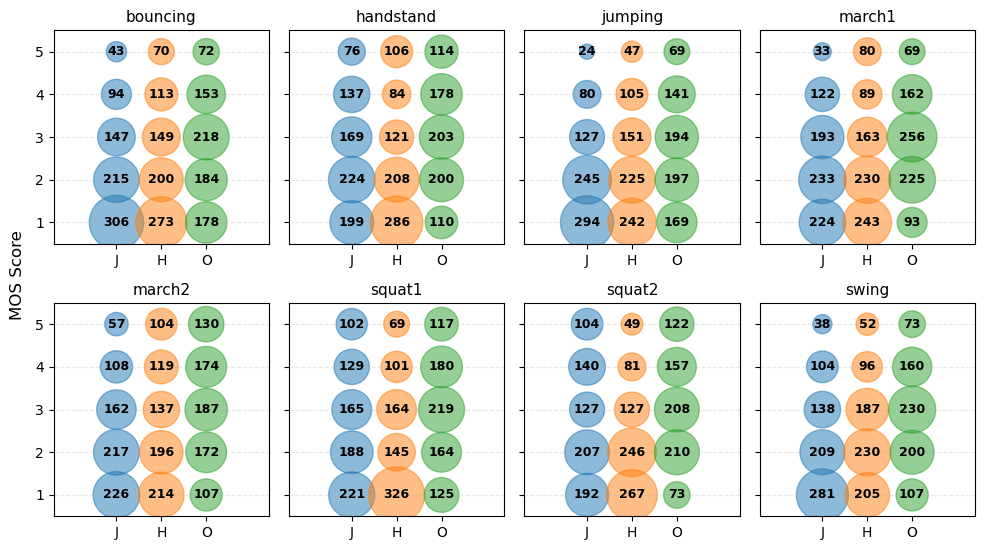}
    \caption{Rating counts for Jitter (J), Holes (H), and Overall (O) over content.}
    \label{fig:range_validity}
\end{figure}

Fig.~\ref{fig:range_validity} summarizes the distribution of ratings for jitter annoyance, hole annoyance, and overall annoyance across stimuli. The results show that all five rating categories are well populated across the full range of artifact severities and combinations,
confirming that the synthesized stimuli span the intended perceptual range. As a result, subsequent analyses of interaction effects and integration behavior are well supported by the data.

\subsubsection{Rating Time and Cognitive Load}
\begin{figure}[h]
    \centering
    \includegraphics[width=0.8\linewidth]{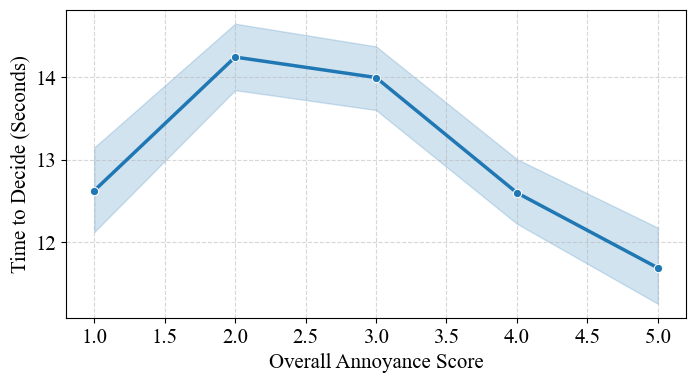}
    \caption{Average time participants take to rate a stimulus.}
    \label{fig:rating_time}
\end{figure}

Fig.~\ref{fig:rating_time} reports the average response time per stimulus as a function of perceived annoyance. Rating times are shortest for stimuli judged as either very low or very high annoyance, and longest for intermediate annoyance levels.

This pattern is consistent with established perceptual decision-making behavior: extreme judgments are made quickly due to high perceptual certainty, while mid-range distortions require greater deliberation. Given that each stimulus is presented for approximately six seconds, participants typically view each clip multiple times before responding, indicating sustained attention and careful evaluation.

Together, these observations suggest that participants remained engaged throughout the experiment and that the collected ratings are reliable.

The complete set of stimuli, subjective quality ratings, and observer demographic metadata are organized into a structured perceptual dataset, which we refer to as the \emph{DynaScan} dataset.

\section{Perceptual Findings}\label{sec:analysis}
This section presents the main perceptual findings. We first analyze the interaction between surface jitter and holes, then examine their joint integration into overall annoyance, and finally study the influence of motion and observer variability. Together, these results reveal systematic, asymmetric, and nonlinear perceptual behaviors.

\subsection{Asymmetric Cross-Artifact Interaction Effects}
\subsubsection{Amplification Effect: Influence of Holes on Perceived Jitter}
\begin{figure}[h]
    \centering
    \includegraphics[width=0.9\linewidth]{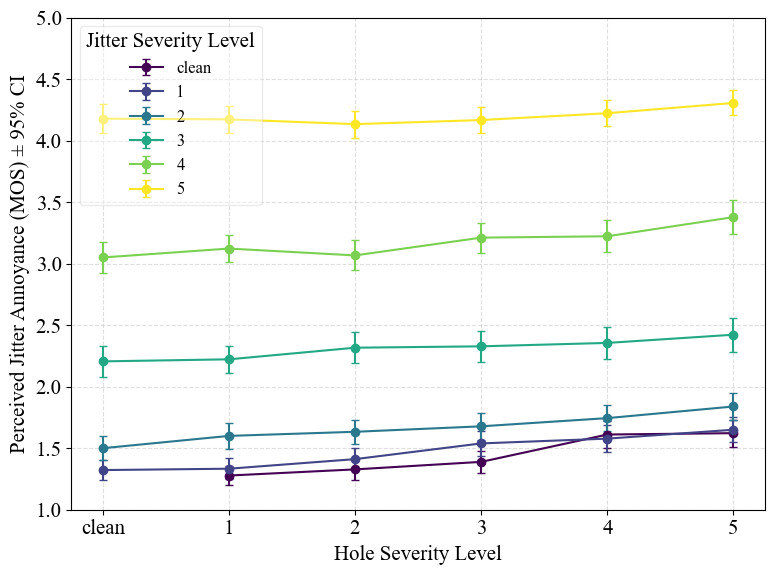}
    \caption{Amplification effect: increasing hole severity amplifies perceived jitter annoyance.}
    \label{fig:cross-amplification}
\end{figure}

To assess whether holes influence the perception of surface jitter, we analyze jitter annoyance ratings while holding the jitter severity constant and varying hole severity. As shown in Fig.~\ref{fig:cross-amplification}, perceived jitter annoyance increases monotonically with hole severity, despite the underlying jitter magnitude remaining unchanged.

These positive slopes indicate an \textbf{amplification effect}, whereby the presence of holes biases observers toward harsher judgments of jitter. An increase in hole severity is associated with a higher perceived annoyance of the same jitter level. In particular, the appearance of holes can visually mimic surface jitter, especially when multiple tiny holes repeatedly appear and disappear across frames. This behavior is most clearly illustrated by the bottom curve in Fig. \ref{fig:cross-amplification}, corresponding to the ``clean" Jitter level, where no surface jitter has been added. Despite the absence of any injected jitter, perceived jitter annoyance increases monotonically with hole severity. 

It can also be seen that the strength of the amplification effect depends on the underlying jitter severity. At low jitter levels, hole severity has a strong influence on perceived jitter annoyance, but this influence diminishes as jitter becomes dominant. This behavior indicates a perceptual saturation effect: once jitter becomes sufficiently salient, additional impairments contribute progressively less to the perceived jitter annoyance.

\subsubsection{Masking Effect: Influence of Jitter on Perceived Holes}
\begin{figure}[h]
    \centering
    \includegraphics[width=0.9\linewidth]{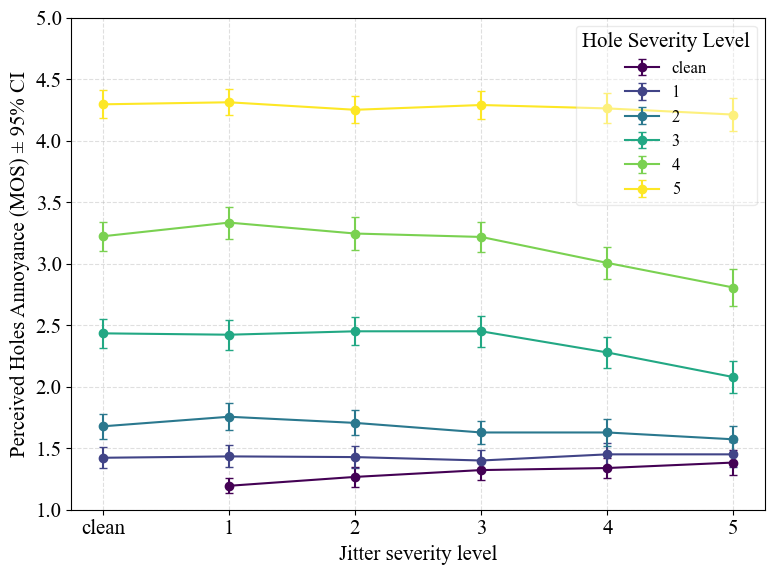}
    \caption{Masking effect: increasing jitter severity reduces perceived hole annoyance at moderate hole levels.}
    \label{fig:cross-masking}
\end{figure}

The reciprocal interaction of how jitter affects hole perception is fundamentally different. When hole severity is held constant and jitter severity is increased (Fig. \ref{fig:cross-masking}), perceived hole annoyance decreases for moderate and high hole levels.

This negative dependency demonstrates a \textbf{masking effect}, in which strong surface instability reduces the visibility or perceptual salience of holes. The effect is not uniform across hole severities. For very low hole levels, masking is negligible, indicating that imperceptible holes cannot be further suppressed. As hole severity increases beyond a perceptual threshold, masking becomes evident and reaches its maximum at intermediate hole levels. For very severe holes, masking weakens again, as large missing regions dominate perception regardless of jitter strength.
This reveals that masking is not a simple monotonic phenomenon, but depends on the prominence of the masked artifact.

\subsection{Motion-Dependent Temporal Masking Across Sequences}
\begin{figure}[h]
    \centering
    \includegraphics[width=0.85\linewidth]{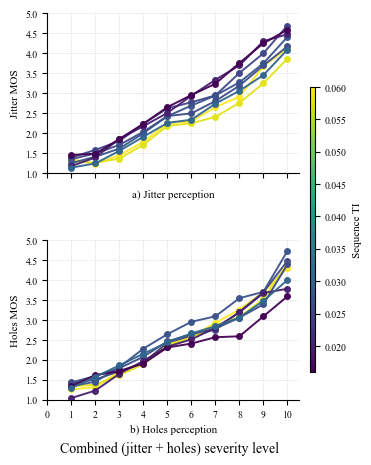}
    \caption{Artifact perceived annoyance. Each curve represents one content sequence, and color encodes motion magnitude.}
    \label{fig:temporal-masking}
\end{figure}

Beyond artifact interactions, perceived annoyance also depends on motion characteristics of the underlying sequence. Fig.~\ref{fig:temporal-masking}a shows that sequences with faster motion (larger TI) consistently receive lower jitter annoyance ratings for the same artifact severity level. This behavior indicates systematic \textbf{motion-dependent temporal masking}, whereby increasing motion strength reduces sensitivity to high-frequency surface instability.
Moreover, the perceptual difference between low- and high-motion sequences increases as jitter severity increases. 

In contrast, hole annoyance (Fig.~\ref{fig:temporal-masking}b) does not exhibit a comparable dependency on motion speed. This asymmetry is consistent with well-established findings in visual perception showing that high-frequency spatiotemporal components are more susceptible to motion masking, whereas low-frequency or spatially structured distortions remain salient \cite{Watson1985}. Surface jitter introduces rapid, high-frequency positional fluctuations that are increasingly masked as motion energy rises, reducing their perceptual visibility. Holes, by contrast, correspond to persistent spatial incompleteness and structural discontinuities, which are dominated by lower-frequency spatial components and therefore remain perceptually salient even under fast motion. 
\subsection{Non-Additive Integration of Multiple Artifacts}
\begin{figure}[h]
    \centering
    \includegraphics[width=\linewidth]{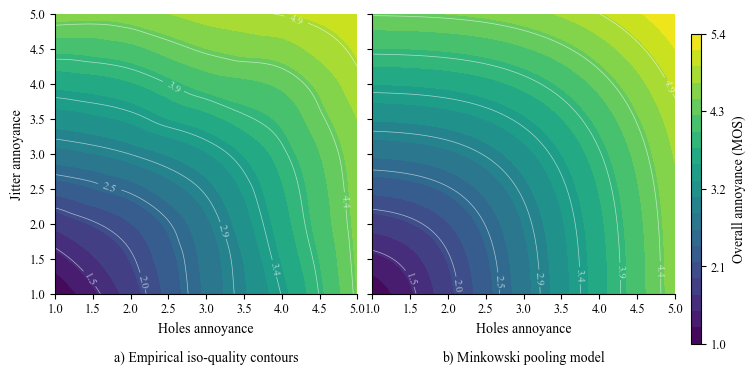}
    \caption{Iso-quality contours in the jitter–holes annoyance plane.}
    \label{fig:contours}
\end{figure}

To understand how jitter and holes jointly contribute to overall visual annoyance, we analyze iso-annoyance contours in the two-dimensional artifact severity space. The resulting contours (Fig. \ref{fig:contours}a) are smooth and exhibit rounded corners.
Such contour geometry is inconsistent with additive integration, which yields linear iso-response boundaries, and with hard max pooling, which produces axis-aligned contours. Instead, it is more consistent with nonlinear pooling mechanisms, such as Minkowski integration, widely adopted in perceptual modeling \cite{Wang2006}. In these formulations, the pooling exponent governs the degree of dominance between components, allowing one artifact to dominate perception without fully suppressing the contribution of the other.

To model this behavior, overall annoyance ($A_O$) is fitted using a Minkowski pooling formulation that combines jitter ($A_J$) and hole ($A_H$) annoyance: $ A_O = \alpha_1(A_J^p+\alpha_2A_H^p)^{1/p}$. 

The fitted model (Fig. \ref{fig:contours}b) accurately reproduces the observed iso-annoyance contours and captures the gradual transition of perceptual dominance between artifacts. The fitted exponent ($p=3.9$) lies well above unity, confirming a strongly nonlinear integration regime intermediate between linear summation ($p=1$) and max pooling ($p=\infty$). The estimated weighting parameters ($\alpha_1=0.91, \alpha_2=0.88$) indicate that jitter and holes contribute comparably to overall annoyance, with a slight dominance of the jitter dimension, indicating that dominance emerging primarily through nonlinear pooling rather than through large differences in linear weights.

These results suggest that additive quality models are insufficient for describing the interaction between holes and jitter in dynamic mesh perception, and that nonlinear pooling models are well-suited to describe their interaction.

\subsection{Inter-Observer Variability and Perceptual Dominance}
\begin{figure}[h]
    \centering
    \includegraphics[width=0.8\linewidth]{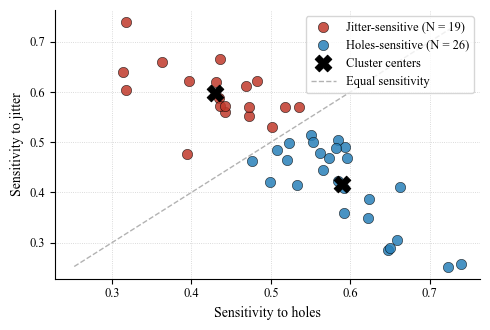}
    \caption{Participant sensitivity profiles to jitter and holes. Sensitivity is quantified as the correlation between overall annoyance ratings and artifact severity levels for each participant.}
    \label{fig:dichotomy}
\end{figure}

Analysis of individual observer responses reveals substantial variability in artifact sensitivity. When correlating overall annoyance ratings with jitter and hole severity separately, participants separate into two dominant groups: jitter-dominant observers and hole-dominant observers (Fig. \ref{fig:dichotomy}).

The observer distribution in Fig. \ref{fig:dichotomy} reveals a clear trade-off in perceptual sensitivity. Instead of concentrating around the equality line, the distribution is oriented approximately perpendicular to it, indicating that increased sensitivity to one artifact systematically coincides with reduced sensitivity to the other. This suggests that observers adopt distinct perceptual strategies, selectively prioritizing either jitter or holes when forming overall quality judgments. Importantly, this structure implies consistent perception rather than inattentive or noisy responses. The dominance pattern is symmetric: observers who are highly sensitive to jitter are no less extreme than those who are highly sensitive to holes. Despite this dichotomy, all observers show non-negligible sensitivity to both artifacts, indicating that no participant is perceptually “blind” to either distortion.

\section{Towards a No-reference Perceptual Metric}

The perceptual findings in Section \ref{sec:analysis} provide evidence that scan-induced artifacts in dynamic meshes are perceived in a nonlinear, asymmetric, and motion-dependent manner. In particular, jitter and holes interact perceptually through amplification and masking effects, integrate via nonlinear integration rather than additive rules, and exhibit distinct sensitivity to motion-induced temporal masking. These findings collectively motivate the design of a no-reference (NR) perceptual metric that encodes such structured perceptual behaviors explicitly, rather than relying solely on geometric distortion measures.

\begin{figure}[h]
    \centering
    \includegraphics[width=\linewidth]{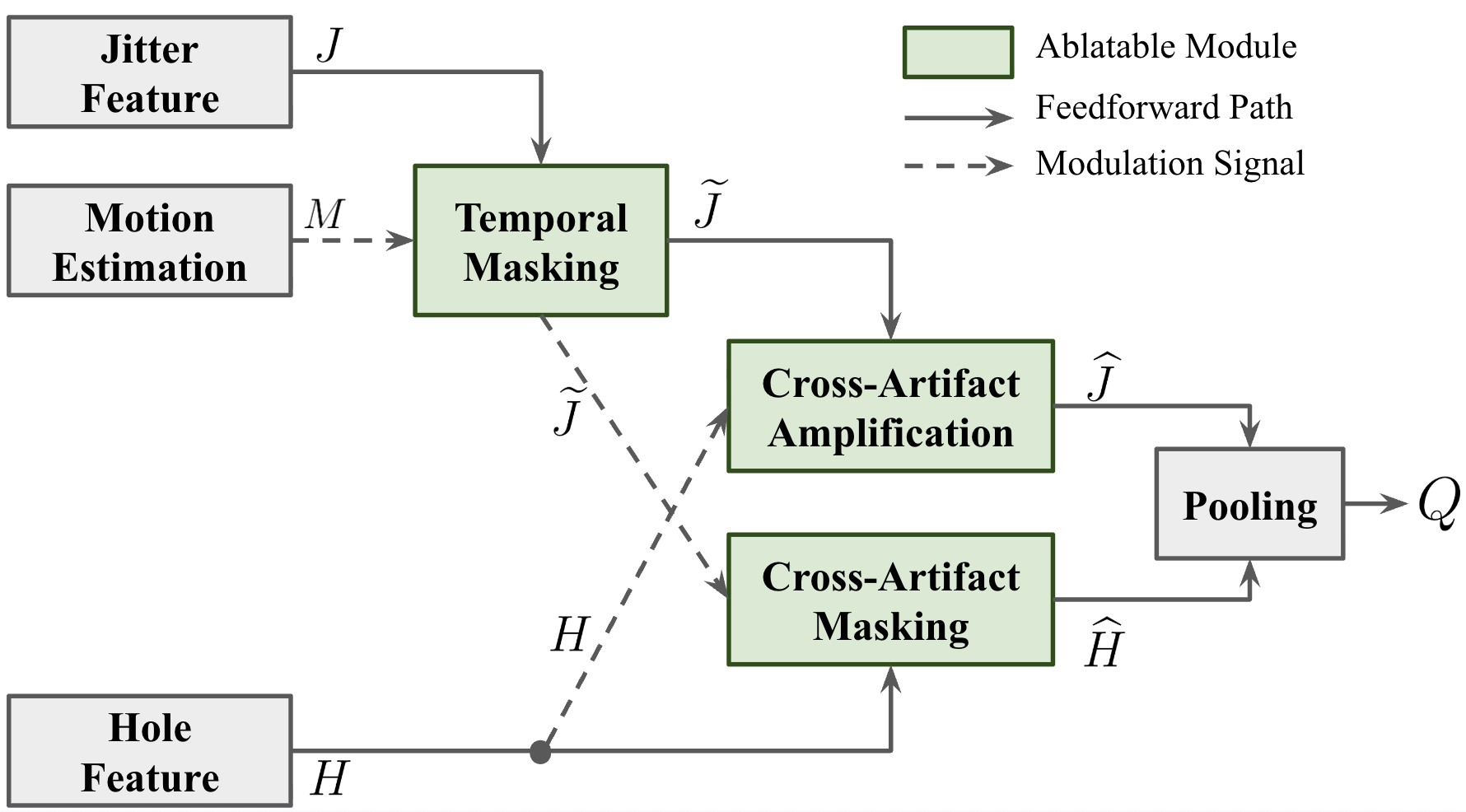}
    \caption{Overview of the PSAM framework.}
    \label{fig:metric-diagram}
\end{figure}

Fig.~\ref{fig:metric-diagram} presents a block diagram of the proposed \emph{Perceptual Scan-induced Artifact Metric (PSAM)} framework. It operates directly on a reconstructed dynamic mesh sequence and produces an overall annoyance score.  First, artifact-specific features are extracted independently to quantify hole severity and jitter severity. Next, the jitter component is modulated by a motion-aware temporal masking stage, reflecting reduced perceptual sensitivity under fast motion. The artifact responses are then passed through a cross-artifact perceptual coupling stage, which models asymmetric amplification and masking interactions between artifacts.  Finally, the adjusted artifact responses are combined to yield an overall annoyance estimate. Each stage corresponds directly to a perceptual phenomenon identified in the subjective study.

A question is whether the perceptual mechanisms, discovered in isolation, retain predictive value when combined in a unified model. In other words, do temporal masking, cross-artifact interactions such as amplification and masking, and nonlinear pooling, provide complementary (additive) explanatory power, or do they exhibit partial redundancy when combined? 
Accordingly, the PSAM framework is intentionally modular; each perceptual mechanism (temporal masking, cross-artifact amplification, and cross-artifact masking) is implemented as an optional nonlinear operator that can be independently enabled or disabled. 
When a module is disabled, it reduces to an identity mapping, and the model collapses to a simpler baseline.

\subsection{Artifact-Specific Feature Selection}
\subsubsection{Hole Severity Feature - Mean Total Hole Area}

Holes are identified as connected components of boundary edges, where a boundary edge is defined as an edge incident to only one triangle. Boundary edges are grouped into boundary loops, each corresponding to a distinct hole. For each frame $t$, the area of a hole is estimated by projecting the vertices of its boundary loop onto a best-fit plane and computing the area of the resulting polygon. Let $A_k(t)$ denote the area of the $k$-th hole at frame $t$. The total hole area at frame $t$ is:
\begin{equation}
    A_\mathrm{holes}(t) = \sum_kA_k(t)
\end{equation}

The hole severity feature is defined as the temporal average of the total hole area normalized by total surface area:
\begin{equation}
    H=\frac{1}{T}\sum_{t=1}^T \frac{A_\mathrm{holes}(t)}{A_\mathrm{surface}(t)}
\end{equation}

\noindent where $T$ is the number of frames, and $A_\mathrm{surface}(t)$ is the total surface area of frame $t$. This feature captures the spatial extent of missing surface regions, which dominates perceptual hole severity.

\subsubsection{Jitter Severity Feature - Dihedral Angle Variance}
Surface jitter manifests as high-frequency geometric irregularity. To quantify this, we compute the dihedral angle between adjacent triangles. For each interior edge $e$ shared by two faces with unit normals $\textbf{n}_1$ and $\textbf{n}_2$, the dihedral angle is defined as:
\begin{equation}
    \theta_e = arccos(\textbf{n}_1 \cdot \textbf{n}_2)
\end{equation}

For each frame $t$, the variance of dihedral angles over all interior edges is computed as:
\begin{equation}
    \sigma_{\theta}^{2}(t) =
\frac{1}{\lvert \mathcal{E}_{\mathrm{int}} \rvert}
\sum_{e \in \mathcal{E}_{\mathrm{int}}}
\left(
\theta_{e} - \bar{\theta}
\right)^{2}
\end{equation}

\noindent where $\mathcal{E}_{\mathrm{int}}$ denotes the set of interior edges and $\bar{\theta}$ is the mean dihedral angle. The jitter severity feature is defined as the temporal average:

\begin{equation}
    J=\frac{1}{T}\sum_{t=1}^{T}\sigma_{\theta}^{2}(t)
\end{equation}

This feature is sensitive to local surface irregularities induced by jitter and largely insensitive to hole extent.

\subsection{Motion-Aware Temporal Masking}
The perceptual study revealed that, for a fixed magnitude of geometric jitter, observers report lower annoyance as object motion increases.

\emph{Motion Estimation:} Reconstructed dynamic meshes frequently lack a fixed or clean topology, which precludes the direct use of the TI formulation in Eq. (\ref{eq:TI}), as that metric relies on explicit vertex correspondences across frames. To address this limitation, we estimate motion using a correspondence-free measure based on a bidirectional Chamfer distance.

Let \(M_t\) and \(M_{t+1}\) denote the meshes at frames \(t\) and \(t+1\), respectively. 
The motion magnitude is defined as:
\begin{equation}
M \;=\; \frac{1}{T-1} \sum_{t=1}^{T-1}
\frac{\operatorname{Chamf}(M_t, M_{t+1})}
{\max\!\left(bb_t,\, bb_{t+1}\right)}
\label{eq:motion}
\end{equation}
where \(bb_t\) and \(bb_{t+1}\) are the diagonal lengths of the axis-aligned bounding boxes of
\(M_t\) and \(M_{t+1}\), respectively. This normalization ensures scale invariance across sequences.

The operator $\operatorname{Chamf}(\cdot,\cdot)$ denotes a variant of the
bidirectional Chamfer distance~\cite{Fan2017}, where the conventional mean is replaced by a trimmed mean to improve robustness to outliers:
\begin{equation}
\begin{aligned}
\operatorname{Chamf}(M_t, M_{t+1})
&= \tfrac{1}{2}\Big[
\operatorname{TM}_{p\%}\!\Big(
\min_{u \in M_{t+1}}
\lVert v-u \rVert_2
\Big)_{v \in M_t} \\
&{}\,+\;
\operatorname{TM}_{p\%}\!\Big(
\min_{v \in M_t}
\lVert u-v \rVert_2
\Big)_{u \in M_{t+1}}
\Big]
\end{aligned}
\label{eq:chamfer}
\end{equation}

\noindent where $\operatorname{TM}_{p\%}$ denotes a trimmed mean computed by discarding the top and bottom $p\%$ of samples from the sorted nearest-neighbor distance distribution. In this work, we use $p=10$ to suppress the influence of outliers.

\emph{Divisive Temporal Masking Model:} Motivated by classical perceptual masking models and our perceptual findings, we model temporal masking of jitter through a normalization formulation where motion attenuates perceived jitter. Specifically, the perceived jitter magnitude $\tilde{J}$ is defined as:

\begin{equation}
    \tilde{J} = \frac{J}{1+\alpha M J}
\end{equation}

\noindent where $J$ denotes the raw jitter magnitude, $M$ is the motion estimate, and $\alpha$ controls the strength of temporal masking. This formulation ensures that attenuation increases jointly with motion strength and artifact severity: for fixed $J$, increasing $M$ reduces $\tilde{J}$ monotonically, and for fixed $M$, higher jitter levels undergo proportionally stronger suppression. 

When Temporal Masking is disabled, $\tilde{J}=J$, and the model reduces to a motion-independent formulation.

\subsection{Cross-Artifact Perceptual Interactions}

The perceptual experiment revealed systematic and asymmetric interactions between holes and jitter: holes can exaggerate the perception of jitter (amplification effect), whereas jitter suppresses the visibility of holes (masking effect). To encode these behaviors, the artifact severities are adjusted through a cross-artifact interaction stage that models both effects as level-dependent nonlinear interactions.

The amplification effect (holes $\rightarrow$ jitter) is modeled as:
\begin{equation}
    \hat{J} = 1 - (1 - \tilde{J})\, e^{-\gamma_a\, H}
\end{equation}
where $\tilde{J}$ is the temporally masked jitter magnitude and $\gamma_a$ controls amplification strength. As $H$ increases, the exponential term decreases, reducing the residual factor $(1-\tilde{J})$ and thereby increasing $\hat{J}$. The amplification is bounded and level-dependent, preventing over-enhancement at high jitter levels while capturing the perceptual exaggeration of jitter in the presence of holes. When the cross-artifact amplification module is disabled (equivalently, $\gamma_a = 0$), the transformation becomes identity, $\hat{J} = \tilde{J}$, and the model no longer models hole-driven exaggeration of jitter saliency.

The masking effect (jitter $\rightarrow$ holes) is modeled as:
\begin{equation}
    \hat{H}
    =
    H\,
    e^ {
    -\gamma_m\,\tilde{J}^{2}\,
    4(H-H^2)}
\end{equation}
where $\gamma_m$ controls masking strength. This formulation enforces three key properties. Suppression increases monotonically with jitter magnitude through $\tilde{J}^2$. The factor $4(H-H^{2}) \in [0,1]$ introduces a mid-range sensitivity profile, peaking at $H=0.5$ and vanishing at the extremes, so masking is strongest when holes are perceptually noticeable but not dominant. The exponential form ensures smooth, bounded attenuation with $\hat{H}=H$ when $\tilde{J}=0$.
Overall, the model captures the experimentally observed nonlinear behavior: jitter selectively suppresses moderately visible holes, with reduced influence at very low or very high hole severities. When the cross-artifact masking module is disabled (i.e., $\gamma_m = 0$), the mapping reduces to $\hat{H} = H$, and the model no longer accounts for jitter-induced suppression of hole visibility.

\subsection{Integration of Overall Annoyance}
The final stage of the pipeline integrates the adjusted severities $\hat{H}$ and $\hat{J}$ into an overall annoyance score. 

Although flexible pooling schemes (e.g., Minkowski integration) can improve performance under some configurations, we observed that additional pooling parameters become unstable once perceptual modules are enabled (see Section \ref{sec:ablation}).
To ensure interpretability and stable generalization, we adopt average pooling:
\begin{equation}
    Q = \frac{1}{2}(\hat{H} + \hat{J})
\end{equation}
This formulation preserves the perceptual structure encoded in preceding stages while avoiding redundant degrees of freedom.

\section{Performance Analysis}

\subsection{Evaluation Protocol}
To evaluate generalization to unseen content while avoiding parameter overfitting, we adopt a nested leave-one-content-out (LOCO) cross-validation framework. Each outer fold holds out all stimuli derived from one sequence for testing, while the remaining sequences are used for parameter estimation. This design explicitly prevents content leakage and reflects the intended deployment scenario in which the metric is applied to previously unseen content.

For each fold, the following metrics are reported:
\begin{itemize}
    \item Pearson Linear Correlation Coefficient (PLCC)
    \item Spearman Rank-Order Correlation Coefficient (SRCC)
    \item Kendall Rank Correlation Coefficient (KRCC)
    \item Root Mean Squared Error (RMSE), computed after applying the train-fitted 4-parameter logistic function (4PL) to the test set.
\end{itemize}

\subsubsection{Nested Cross-Validation Procedure}
For each outer LOCO fold, model parameters are estimated exclusively on the training data using an inner cross-validation loop. Specifically, a 5-fold inner cross-validation, stratified by distortion condition, is employed to optimize the model parameters by maximizing agreement with subjective MOS. The optimization objective is defined as

\begin{equation}
\arg\max_{\theta} \;
\left(
\mathrm{PLCC}(Q_{\theta}, \mathrm{MOS})
+ \mathrm{SRCC}(Q_{\theta}, \mathrm{MOS})
\right)
\end{equation}

\noindent where $Q_{\theta}$ denotes the predicted overall quality score produced by the proposed metric, and $\theta$ is the set of parameters to be estimated. We use the L-BFGS-B algorithm to optimize the parameters, performing 5 random restarts with different initial points to reduce the risk of convergence to local minima.

\subsubsection{Stratification}
The inner cross-validation folds are stratified according to distortion conditions (holes-only, jitter-only, and combined distortions). This stratification mitigates the inherent imbalance in the dataset, where combined distortions are more prevalent than single-artifact cases, and ensures that each inner fold contains representative samples of all distortion conditions. This prevents the parameter optimization from being dominated by combined-distortion stimuli and improves the stability of the learned parameters.

\subsubsection{Feature Normalization}
The hole severity $H$, jitter severity $J$, and motion magnitude $M$ are derived from different geometric quantities and therefore may exhibit substantially different numerical ranges. Directly combining features with mismatched scales can bias parameter optimization and affect numerical stability. To ensure balanced contribution of each feature during model estimation, we apply linear min--max normalization.
To prevent information leakage, normalization parameters are computed exclusively from the training portion of each outer LOCO fold and subsequently applied to the corresponding test split.
Min--max normalization preserves monotonic relationships and relative ordering within each fold. As a result, the normalization does not alter correlation structure or distort linear relationships between features and subjective scores, while improving numerical conditioning during parameter optimization.

\subsection{Systematic Module Combination Analysis}
\label{sec:ablation}

PSAM's modular design allows each perceptual mechanism to be treated as a hypothesis-driven operator whose contribution can be evaluated independently and in combination. Rather than viewing the model as a fixed pipeline, we interpret it as a design space defined by temporal masking (TM), cross-artifact amplification (CA), cross-artifact masking (CM), and pooling strategy. The central question is whether these mechanisms provide complementary explanatory power when jointly instantiated, or whether they exhibit partial redundancy by modulating a shared saliency dimension. We therefore evaluate all structurally meaningful combinations under the same nested LOCO protocol. This analysis provides a controlled comparison of model variants and enables empirical validation of the perceptual mechanisms inferred from the subjective study. Results are summarized in Table~\ref{tab:avg_ablation}.

\begin{table*}[h]
\centering
\caption{LOCO performance (Mean $\pm$ STD across folds) comparison of ablation models grouped by structural tier.}
\label{tab:avg_ablation}
\small
\setlength{\tabcolsep}{5pt}
\renewcommand{\arraystretch}{1}

\begin{tabular}{llccccc}
\toprule
\textbf{} & \textbf{Model} & \textbf{\# Params} 
& \textbf{PLCC} $\uparrow$ 
& \textbf{SRCC} $\uparrow$ 
& \textbf{KRCC} $\uparrow$ 
& \textbf{RMSE} $\downarrow$ \\
\midrule

\multicolumn{7}{l}{\textbf{Single-Artifact Baselines}} \\
\addlinespace[2pt]
\textit{M1} & Jitter Only 
& 0 
& 0.6007 $\pm$ 0.0630 
& 0.4224 $\pm$ 0.1174 
& 0.2834 $\pm$ 0.0874 
& 0.7878 $\pm$ 0.0668 \\

\textit{M2} & Hole Only 
& 0 
& 0.6479 $\pm$ 0.0645 
& 0.5414 $\pm$ 0.0785 
& 0.3840 $\pm$ 0.0641 
& 0.7590 $\pm$ 0.0903 \\

\addlinespace[6pt]
\multicolumn{7}{l}{\textbf{Fusion Baselines}} \\
\addlinespace[2pt]

\textit{M3} & Average Pooling (AVG) 
& 0 
& 0.9199 $\pm$ 0.0256 
& 0.9679 $\pm$ 0.0153 
& 0.8661 $\pm$ 0.0421 
& 0.3486 $\pm$ 0.0688 \\

\textit{M4} & Minkowski Pooling (MINK) 
& 2 
& 0.9271 $\pm$ 0.0205 
& 0.9710 $\pm$ 0.0134 
& 0.8754 $\pm$ 0.0383 
& 0.3412 $\pm$ 0.0638 \\

\addlinespace[6pt]
\multicolumn{7}{l}{\textbf{Perceptual Mechanisms}} \\
\addlinespace[2pt]

\textit{M5} & AVG + TM
& 1 
& 0.9231 $\pm$ 0.0198 
& 0.9723 $\pm$ 0.0126 
& 0.8780 $\pm$ 0.0350 
& 0.3203 $\pm$ 0.0608 \\

\textit{M6} & AVG + CA (\textbf{PSAM-A})
& 1 
& 0.9339 $\pm$ 0.0151 
& 0.9759 $\pm$ 0.0109 
& 0.8881 $\pm$ 0.0305 
& 0.3331 $\pm$ 0.0645 \\

\textit{M7} & AVG + CM
& 1 
& 0.9316 $\pm$ 0.0216 
& 0.9665 $\pm$ 0.0162 
& 0.8628 $\pm$ 0.0439 
& 0.3504 $\pm$ 0.0688 \\

\addlinespace[6pt]
\multicolumn{7}{l}{\textbf{Joint Perceptual Models}} \\
\addlinespace[2pt]

\textit{M8} & AVG + CA + CM (\textbf{PSAM-I})
& 2 
& 0.9383 $\pm$ 0.0142
& 0.9744 $\pm$ 0.0124
& 0.8860 $\pm$ 0.0348
& 0.3355 $\pm$ 0.0639 \\

\textit{M9} & AVG + TM + CA + CM 
& 3 
& 0.9352 $\pm$ 0.0162 
& 0.9709 $\pm$ 0.0121 
& 0.8742 $\pm$ 0.0337 
& 0.3305 $\pm$ 0.0665 \\

\textit{M10} & MINK + TM + CA + CM 
& 5 
& 0.9600 $\pm$ 0.0102 
& 0.9723 $\pm$ 0.0099 
& 0.8784 $\pm$ 0.0275 
& 0.3105 $\pm$ 0.0700 \\

\bottomrule
\end{tabular}
\end{table*}

\paragraph{Baseline Behavior}
Single-artifact models (hole-only and jitter-only) yield substantially lower correlations, confirming that overall annoyance cannot be explained by either artifact dimension in isolation. 
Simple average pooling (AVG) of hole and jitter severities provides a strong baseline (PLCC = 0.9199), indicating that both artifact types contribute meaningfully to perceived annoyance.  Replacing naive additive fusion with Minkowski pooling of hole and jitter severities,
\begin{equation}
    Q = (w\hat{H}^p + (1-w)\hat{J}^p)^{1/p}
\end{equation}
improves PLCC (0.9271), indicating that without perceptual modules, nonlinear integration is beneficial when combining raw artifact magnitudes. 

\paragraph{Effect of Individual Perceptual Modules}
Each perceptual module introduces a hypothesis-driven structural modification to the baseline.
TM improves rank correlations and reduces RMSE, demonstrating that motion-dependent attenuation increases monotonic alignment with subjective judgments. 
CA produces the largest individual improvement over the baseline, confirming that hole-induced exaggeration of jitter is a dominant perceptual effect.
CM also improves correlation relative to baseline, validating the experimentally observed suppression of moderately visible holes by jitter.
These results confirm that each mechanism independently captures a perceptually meaningful component.

\paragraph{Cross-Artifact Interaction Models}

When CA and CM are combined (M8), the model achieves strong and consistent performance under LOCO (PLCC = 0.9383 $\pm$ 0.0142), slightly outperforming all single-mechanism variants. Importantly, both interaction parameters remain stable across LOCO folds ($\gamma_a = 0.30 \pm 0.06$, $\gamma_m = 0.63 \pm 0.18$), indicating that amplification and masking represent complementary and consistently estimable perceptual contributions.

Although the correlation improvement over CA alone (M6) is modest relative to fold variability, M8 is the minimal formulation that jointly instantiates both asymmetric cross-artifact mechanisms observed in the subjective study. We therefore adopt M8 as the primary model, denoted \textbf{PSAM-I}, as it achieves the best trade-off between predictive strength, parameter stability, and perceptual interpretability.

For applications prioritizing reduced complexity, M6 (AVG + CA), denoted \textbf{PSAM-A}, provides a lightweight alternative. It achieves comparable predictive accuracy with a single parameter.

\paragraph{Interaction Between Temporal Masking and Cross-Artifact Interactions}
Adding temporal masking to the cross-interaction model (M9) yields comparable overall performance but does not improve correlations relative to M8, while modestly improving RMSE. Under joint modeling, the amplification parameter magnitude collapses ($\gamma_a = 0.02 \pm 0.06$), while cross-artifact masking and temporal masking remain relatively stable ($\gamma_m = 1.40 \pm 0.24$, $\alpha = 0.59 \pm 0.25$).

This behavior indicates partial redundancy: both temporal masking and cross-artifact amplification modulate jitter dimension, and when modeled simultaneously, their effects overlap. This provides confirmation that perceptual processing is layered and partially overlapping rather than strictly additive, especially when applied to the same feature dimension.

We therefore interpret M9 not as a superior predictive model, but as evidence that motion-dependent masking and cross-artifact amplification operate along closely related perceptual dimensions. Temporal masking is thus validated as an independent perceptual effect (M5), but its separability from amplification is limited under joint parameterization.

\paragraph{Flexible Pooling as an Upper Bound}
Replacing average pooling with Minkowski integration (M10) improves correlation (PLCC = 0.96). 
However, the pooling parameters consistently approach optimization bounds and exhibit substantial fold-to-fold variability.
This suggests that the added flexibility may reflect over-parameterization rather than a robust perceptual effect.
Given the goal of constructing a compact and interpretable perceptual model, we therefore prioritize stable formulations over marginal gains obtained through unstable parameters.

Taken together, these results demonstrate that the proposed perceptual mechanisms provide structured and interpretable improvements over raw objective feature fusion. PSAM-I represents the minimal model that jointly captures both asymmetric cross-artifact effects with stable parameter estimates. PSAM-A serves as a lightweight variant with comparable accuracy and reduced complexity. 

This layered analysis confirms that perceptual integration in dynamic meshes is partially complementary and partially overlapping, and underscores the importance of stability and interpretability when translating perceptual findings into predictive models.

\subsection{Fold-wise Performance of PSAM-I}

\begin{figure}[h]
    \centering
    \includegraphics[width=\linewidth]{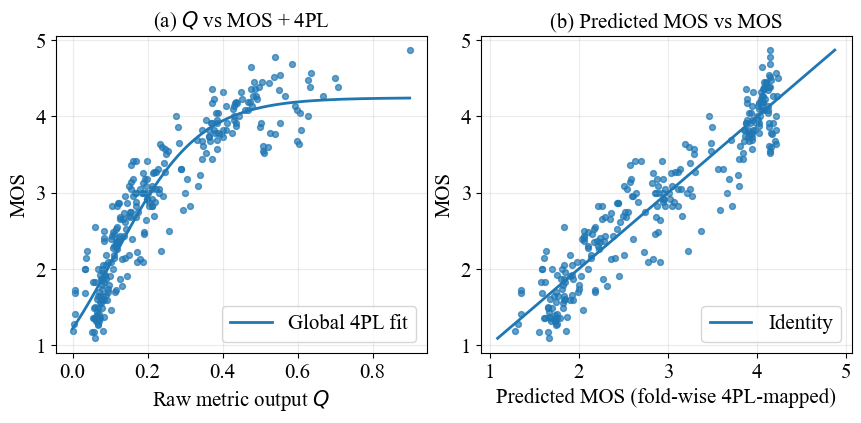}
    \caption{
    Relationship between predicted quality and subjective MOS under nested LOCO cross-validation.
    (a) Raw metric output $Q$ versus MOS with a 4PL curve fitted for visualization.
    (b) Out-of-fold predicted MOS after applying train-fitted logistic mapping, plotted against subjective MOS.
    }
    \label{fig:dual_psychometric}
\end{figure}

Table \ref{tab:overall_performance} reports the fold-wise predictive performance of the PSAM-I metric. The model achieves strong and consistent correlation with subjective overall annoyance scores across all folds, demonstrating reliable generalization under the LOCO protocol. Figure~\ref{fig:dual_psychometric} provides a visualization of predictive behavior.
Panel (a) illustrates the monotonic relationship between the raw metric output and subjective scores, while panel (b) reflects the final evaluation protocol by plotting out-of-fold mapped predictions against MOS.
The tight clustering around the identity line confirms strong agreement and stable calibration across unseen content.

\begin{table}[t]
\centering
\caption{Per-sequence performance of the proposed NR model (PSAM-I) under nested LOCO cross-validation.}
\label{tab:overall_performance}
\small
\setlength{\tabcolsep}{8pt}
\begin{tabular}{l
                S[table-format=1.4]
                S[table-format=1.4]
                S[table-format=1.4]
                S[table-format=1.4]}
\toprule
\textbf{Sequence} &
{\textbf{PLCC} $\uparrow$} &
{\textbf{SRCC} $\uparrow$} &
{\textbf{KRCC} $\uparrow$} &
{\textbf{RMSE} $\downarrow$} \\
\midrule
bouncing  & 0.9380 & 0.9669 & 0.8545 & 0.3010 \\
handstand & 0.9415 & 0.9689 & 0.8622 & 0.3877 \\
jumping   & 0.9510 & 0.9650 & 0.8603 & 0.4632 \\
march1    & 0.9423 & 0.9875 & 0.9240 & 0.2724 \\
march2    & 0.9069 & 0.9529 & 0.8386 & 0.3833 \\
squat1    & 0.9307 & 0.9826 & 0.9080 & 0.2821 \\
squat2    & 0.9381 & 0.9926 & 0.9418 & 0.2830 \\
swing     & 0.9580 & 0.9788 & 0.8985 & 0.3115 \\
\midrule
\textbf{Mean} & \bfseries \textbf{0.9383} & \bfseries \textbf{0.9744} & \bfseries \textbf{0.8860} & \bfseries \textbf{0.3355} \\
Std  & \bfseries 0.0142 & \bfseries 0.0124 & \bfseries 0.0348 & \bfseries 0.0639 \\
\bottomrule
\end{tabular}
\end{table}

\subsection{Validation of Designed Features}

We validate the proposed hole and jitter descriptors by analyzing their relationship with subjective artifact-specific scores. Table~\ref{tab:feature_validation} reports the correlation between each feature and the corresponding MOS. The hole feature exhibits a strong correlation with MOS$_{\mathrm{holes}}$ (PLCC = 0.899, SRCC = 0.935), while the jitter feature shows similarly strong correlation with MOS$_{\mathrm{jitter}}$ (PLCC = 0.906).
In contrast, cross-artifact correlations are negligible: the hole feature is weakly correlated with MOS$_{\mathrm{jitter}}$, and the jitter feature shows low correlation with MOS$_{\mathrm{holes}}$. Moreover, the two features themselves exhibit low mutual correlation, indicating limited redundancy. These results demonstrate that the proposed descriptors are artifact-specific, minimally redundant, and aligned with the perceptual dimensions measured in the subjective study. 
The low cross-correlation confirms that hole and jitter severities occupy distinct perceptual axes, justifying their separate treatment in the subsequent perceptual modeling stages.

\begin{table}[h]
\centering
\caption{Correlation analysis validating the specificity and complementarity of the designed hole and jitter features.}
\small
\setlength{\tabcolsep}{6pt}
\begin{tabular}{lccc}
\hline
\textbf{Feature Pair} & \textbf{PLCC} & \textbf{SRCC} & \textbf{KRCC} \\
\hline
Hole feature vs. MOS$_\text{holes}$   & 0.899 & 0.935 & 0.777 \\
Jitter feature vs. MOS$_\text{jitter}$& 0.906 & 0.853 & 0.671 \\
Hole feature vs. MOS$_\text{jitter}$  & 0.043 & 0.081 & 0.056 \\
Jitter feature vs. MOS$_\text{holes}$ & -0.135 & -0.240 & -0.164 \\
Hole feature vs. Jitter feature       & -0.087 & -0.239 & -0.174 \\
\hline
\end{tabular}
\label{tab:feature_validation}
\end{table}

\begin{figure}[h]
    \centering
    \includegraphics[width=\linewidth]{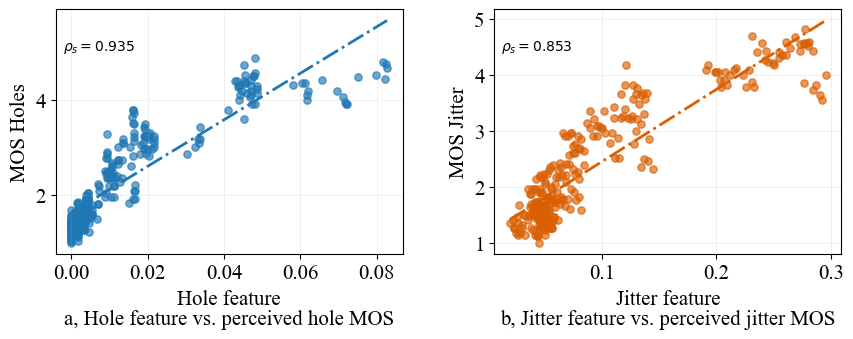}
    \caption{Relationship between proposed features and subjective artifact scores.}
    \label{fig:feature-corr}
\end{figure}

Fig.~\ref{fig:feature-corr} further illustrates these findings through scatter plots, where clear monotonic trends are observed between each feature and its corresponding perceptual score. Together, these results confirm that the designed features faithfully capture distinct perceptual characteristics of hole and jitter artifacts.

\subsection{Comparison with Full-Reference Mesh Quality Metrics}

To contextualize the proposed NR metric, we compare it with representative full-reference (FR) metrics that require access to reference meshes.
Specifically, we consider MSDM2 \cite{Lavou2011}, as well as widely adopted distortion measures from the MPEG dynamic mesh evaluation framework \cite{mmetric}. These include D1 (point-to-point distance), D2 (point-to-plane distance), and GEO-PSNR, which is derived from the mean squared D1 and D2 errors. We also evaluate Hausdorff-based variants, D1-Hausdorff and D2-Hausdorff, where distances are computed using the Hausdorff metric.

\begin{table}[h]
\centering
\caption{
Comparison against representative FR mesh quality metrics.
}
\small
\setlength{\tabcolsep}{7pt}
\renewcommand{\arraystretch}{1.15}
\begin{tabular}{lccc}
\hline
\textbf{Metric} & \textbf{Type} & \textbf{PLCC} & \textbf{SRCC} \\
\hline
D2 & FR & $0.9122$ & $0.9246$ \\
D1 & FR & $0.8800$ & $0.8980$ \\
GEO PSNR & FR & $0.8063$ & $0.8166$ \\
MSDM2 & FR & $0.7008$ & $0.6956$ \\
D2-Hausdorff & FR & $0.6421$ & $0.6622$ \\
D1-Hausdorff & FR & $0.5976$ & $0.6048$ \\
\hline
\textbf{PSAM-I (proposed)} & \textbf{NR} & $\mathbf{0.9383}$ & $\mathbf{0.9744}$ \\
\hline
\end{tabular}

\label{tab:fr_comparison}
\end{table}

Table \ref{tab:fr_comparison} shows that even with access to the ground-truth mesh, the performance of the FR metrics under the \textit{DynaScan} dataset still falls short compared to that of the proposed NR metric, which operates directly on the reconstructed mesh.
The best-performing FR metric (D2) achieves PLCC of 0.9122, while the proposed NR metric achieves PLCC of 0.9383. 
The performance gap between the proposed NR metric and existing FR metrics should be interpreted in light of their design assumptions. 
Most full-reference mesh quality metrics were originally developed for compression and simplification scenarios, where the reference geometry is generally assumed to be complete. 
These metrics primarily measure geometric deviation between corresponding surfaces, such as point-to-point distance, Hausdorff distance, or curvature distortion.

In contrast, the reconstruction artifacts studied in this work include topological defects such as holes. Missing surface regions fundamentally violate the assumption of relative correspondence to a complete reference mesh. 
As a result, geometric deviation metrics may underestimate the perceptual severity of holes, particularly when the missing regions occupy limited surface area but are perceptually salient.

Our analysis confirms this behavior. 
Metrics based on surface distance or curvature variation exhibit strong correlation with jitter-related annoyance but substantially weaker correlation with hole-related annoyance. 
This distortion-dependent sensitivity suggests that these metrics are effective at capturing geometric noise but do not explicitly model missing structure.

The PSAM-I metric, in contrast, includes an explicit hole severity feature that directly quantifies surface loss. 
This structural inclusion enables the model to account for perceptual sensitivity to topological defects that are not well represented by geometric deviation alone. 
The improved overall correlation therefore arises not merely from nonlinear modeling, but from incorporating artifact representations aligned with the types of distortions present in dynamic reconstructions.

\section{Conclusions and Future Work}
This paper presents a systematic perceptual characterization of surface jitter and missing geometry, two types of scan-induced artifacts in dynamic 3D mesh reconstructions. To enable systematic experimentation, we introduced a physically grounded artifact synthesis framework in which hole formation is driven by multi-view reconstruction uncertainty. This design produced realistic stimuli and enabled a comprehensive subjective study spanning isolated artifacts and their interactions. The resulting stimuli and perceptual ratings are consolidated in the \textit{DynaScan} dataset, establishing a controlled benchmark for studying scan-induced artifacts.

The subjective results reveal that reconstruction artifacts in dynamic meshes are perceived through structured, non-additive mechanisms rather than independent or purely geometric error accumulation. We found (i) \emph{asymmetric cross-artifact interactions}: holes amplify perceived jitter (amplification effect), while jitter suppresses hole visibility (masking effect); (ii) \emph{motion-dependent temporal masking} that selectively attenuates perceived jitter under faster motion, with comparatively limited impact on hole perception; and (iii) \emph{nonlinear integration} of multiple artifacts, evidenced by smooth iso-annoyance contours that are inconsistent with linear summation or hard max pooling.

Guided by these findings, we proposed a compact NR perceptual quality framework that encodes the identified mechanisms. 
We examined whether the perceptual mechanisms retain additive explanatory power when combined in a unified NR quality model. The results show that asymmetric cross-artifact interactions provide stable and complementary predictive benefit beyond linear fusion, while temporal masking exhibits partial redundancy with cross-artifact amplification, revealing overlapping saliency mechanisms. The adopted \textit{PSAM-I} metric achieves a strong correlation with subjective overall annoyance and outperforms representative full-reference mesh metrics on this reconstruction-focused benchmark. 

Overall, this work advances the perceptual understanding of dynamic mesh reconstructions and provides (i) a methodology for generating physically plausible, controllable scan-induced artifacts, (ii) an extensive perceptual and structural analysis that uncovers asymmetric cross-artifact interactions, motion-dependent masking, and non-additive integration, and systematically evaluates whether these mechanisms retain additive explanatory power when jointly instantiated, and (iii) a perceptually grounded NR quality assessment with strong correlation with human perception. We expect these contributions to support future work in perceptually optimized reconstruction, filtering, and compression.

Several directions naturally extend this work.
First, while this study focused on surface jitter and holes, real-world reconstructions may exhibit additional artifact types, such as temporal topology changes, surface swimming, or texture instability. Extending the perceptual framework to incorporate these artifacts and to study higher-order interaction effects would enrich the understanding of dynamic mesh perception.

Second, the ablation analysis revealed potential partial overlap between temporal masking and cross-artifact amplification modules. Future work could explore hierarchical or factorized formulations that explicitly model shared underlying perceptual channels rather than treating mechanisms as independent operators.

Third, the current framework models average observer behavior, but as dominance weighting between artifact types may vary across individuals, incorporating adaptive or observer-dependent weighting schemes could improve predictive robustness and enable personalized perceptual quality assessment.

Finally, the perceptual findings reported in this work can inform algorithm design beyond quality assessment. Reconstruction, denoising, and compression pipelines may benefit from explicitly accounting for asymmetric artifact interactions and motion-dependent masking. By integrating perceptual sensitivity models into optimization objectives, future systems can better allocate distortion where it is perceptually less salient while suppressing artifacts that disproportionately degrade visual experience.

To support reproducibility, the \textit{DynaScan} dataset, artifact-generation codebase, and the PSAM framework with associated metrics will be made publicly available at: \textit{[URL to be added upon publication]}.

\bibliographystyle{IEEEtran}
\bibliography{egbib}

\vfill

\end{document}